\documentclass[aps,prl,twocolumn,amsmath,amssymb,graphicx,superscriptaddress]{revtex4-1}

\usepackage{bm}
\usepackage{graphicx}
\usepackage{amsthm}
\usepackage{notes2bib}
\usepackage{amsmath}
\usepackage{amssymb}
\usepackage{url}
\usepackage{enumerate}
\usepackage{epstopdf}
\usepackage{color}
\usepackage{mathtools} 
\usepackage{setspace}
\usepackage{appendix}
\usepackage[normalem]{ulem}
\makeatletter
\let\MYcaption\@makecaption
\makeatother
\usepackage{subcaption}
\makeatletter
\let\@makecaption\MYcaption
\makeatother

\usepackage[usenames,dvipsnames,svgnames,table]{xcolor}
\usepackage{color}

\newcommand{\eps}{\varepsilon}
\newcommand{\etal}{\textit{et al}.}

\newcommand{\eg}{\textit{e}.\textit{g}. }

\begin{document}
\title{A laminar chaotic saddle within a turbulent attractor}

\author{Hibiki Kato}
\affiliation{Faculty of Commerce and Management, Hitotsubashi University, Tokyo 186-8601, Japan}
\author{Miki U Kobayashi}
\affiliation{Faculty of Economics, Rissho University, Tokyo 141-8602, Japan}
\author{Yoshitaka Saiki}
\affiliation{Graduate School of Business Administration, Hitotsubashi University, Tokyo 186-8601, Japan}
\author{James A. Yorke}
\affiliation{University of Maryland, College Park, Maryland 20742, USA}

\begin{abstract}
    Intermittent switchings between weakly chaotic (laminar) and strongly chaotic (bursty) states are often observed in systems with high-dimensional chaotic attractors, such as fluid turbulence. They differ from the intermittency of a low-dimensional system accompanied by the stability change of a fixed point or a periodic orbit in that the intermittency of a high-dimensional system tends to appear in a wide range of parameters.
    This paper considers a case where the skeleton of a laminar state $L$ exists as a proper chaotic subset $S$ of a chaotic attractor $X$, that is, $S\ \subsetneq\ X$.   
    We characterize such a laminar state $L$ by a chaotic saddle $S$, which is densely filled with periodic orbits of different numbers of unstable directions.
    This study demonstrates the presence of chaotic saddles underlying intermittency in fluid turbulence and phase synchronization.
    Furthermore, we confirm that chaotic saddles persist for a wide range of parameters. Also, a kind of phase synchronization turns out to occur in the turbulent model. 
\end{abstract}
\date{\today}
\maketitle

\section{I. Introduction} 
    \noindent In nonlinear phenomena, we often observe intermittent dynamics in which weakly chaotic states (laminar) and strongly chaotic states (bursty) alternately appear. 
    Various types of intermittency exist, such as the Pomeau--Manneville intermittency~\cite{pomeau1980}, crisis induced intermittency~\cite{grebogi1987a}, on--off intermittency~\cite{fujisaka1985,pecora1990,platt1993,glendinning2001,czajkowski2024}, chaotic itinerancy~\cite{kaneko2003}, and intermittency in fluid dynamics~\cite{frisch1995}.
    This paper provides a unifying dynamical system approach to characterizing laminar states in high-dimensional systems, including fluid turbulence and coupled chaotic oscillators that are asymmetric to each other.  
    Accordingly, we consider a case where the skeleton of a laminar state $L$ exists as a proper chaotic subset $S$ of a chaotic attractor $X$, that is, $S\subsetneq X$. 
    We recognize $S$ as a (laminar) chaotic saddle and expect that the laminar state $L$, on which typically long-lived regular behavior is observed,  
    corresponds to the localized chaotic saddle $S$.
    We continue following the laminar chaotic saddle even for parameter values where $L$ is short-lived. 

    A chaotic attractor and a chaotic saddle are both chaotic invariant sets~\cite{lai2011transient},  each characterized by the presence of an infinite number of unstable periodic orbits embedded within it. 
    However, unlike a chaotic attractor that attracts arbitrary neighborhoods, a chaotic saddle has non-attracting neighborhoods. 
    Because of its instability, a chaotic saddle is not captured in the time-positive direction, though it has an important role in the dynamics~\cite{lai1999, medeiros2021, rempel2005}. 
    Our research extends the study of chaotic dynamics through the analysis of the core, saddle invariant set. Specifically, the role of chaotic saddle is greater than that of typical saddle invariant sets such as an unstable fixed point~\cite{nishiura1999} and an unstable periodic orbit~\cite{kawahara2001}. 
    We use a specific method called the stagger-and-step method to detect a chaotic saddle~\cite{sweet2001}.

    In this paper, we numerically show the existence of a hetero-chaotic saddle underlying fluid turbulence and phase synchronization, showing intermittency. 
    Hetero-chaotic saddle denotes a chaotic saddle containing multiple periodic orbits with different unstable dimensions~\cite{hyper-chaos, saiki2021}. 
    We consider that the maximal chaotic saddle is identified in the laminar state. Furthermore, we show that the chaotic saddle exists for a wide range of parameters, including the case where the laminar state is invisible.
    Our research is in the same line of research on chaotic dynamics focusing on a saddle invariant set, such as an unstable fixed point~\cite{nishiura1999} and an unstable periodic orbit~\cite{kawahara2001}. Because of the size of the chaotic saddle, its role is bigger than that of a saddle fixed point or a saddle periodic orbit in the dynamics. The situation is similar to Grebogi, Ott, and Yorke's analysis~\cite{grebogi1983c, grebogi1985a}, which observes the super persistent transient behavior before escaping to infinity, whose skeleton is the fractal-shaped chaotic saddle. 

    In fluid dynamics, intermittency is typically characterized by statistical properties rather than the dynamics.
    By considering invariant sets, we aim to characterize intermittency from a dynamical systems perspective. 
    Furthermore, in the study of intermittency, the definition of laminar regions is based on arbitrary criteria.
    By characterizing laminar $L$ regions using invariant sets $S$, we suggest a more rigorous definition.

    In Section II, we analyze the intermittency in the model that mimics fluid turbulence.
    In Section III, we analyze the coupled R\"ossler model.
    Finally, in Section IV, a summary and discussion are presented.

\section{II. Fluid turbulence model}\label{sec:fluid}
    In this section, we show the existence of a hetero-chaotic saddle corresponding to the laminar flow inside the turbulent attractor and that such a chaotic saddle persists for a wide range of parameters, including the parameter where laminar behavior does not appear frequently.
    
    Shell models of fluid turbulence have been introduced to understand the chaotic properties of fully developed turbulence. 
    We investigate the chaotic properties of switching between weak and strongly chaotic regimes by employing the Gledzer--Ohkitani--Yamada (GOY) shell model~\cite{yamada1987}. The GOY shell model is a system of ordinary differential equations of complex variables $u_j$ described as follows: 
    \begin{align}
    &\left(\frac{d}{dt}+\nu k^{2}_{j}\right)u_j=i [A_j u_{j+1} u_{j+2} +B_j u_{j-1}u_{j+1} \nonumber\\
    &+C_j u_{j-1} u_{j-2}]^{*}+f\delta_{j,1} ~ (j=1,\cdots,N), 
    \label{eq:shell-model}
    \end{align}
    where
    $A_N=A_{N-1}=B_N=B_1=C_1=C_2=0$,\ for other $j$'s $A_j=k_j,\ B_j=-\beta k_j,\ C_j=(\beta-1) k_{j-2},$ and 
    $f=5 (1+i) 10^{-3}$ is an external force, and 
    $k_j=2^{j-4}$.
    The model mimics the Fourier spectral Navier--Stokes Equation with periodic boundary conditions 
    and can show the Kolmogorov spectrum and intermittency. In our investigation, we choose $N=14$ , $\nu=1.8\times10^{-4}$ and various $\beta$'s. 

    As $\beta$ increases from the parameter where the dynamics is weakly chaotic, intermittent behavior occurs at the critical parameter $\beta^*\ \approx\ 0.41616$.
    At this parameter, the system undergoes an interior-crisis~\cite{kuramoto-shivashinsky}. 
    Around $\beta=0.4162$, the time evolution of the shell model is a mixture of the quiescent phase (e.g. from $t=2500$ to $t=5000$ in Fig.~1a) and bursty phase (e.g., from $t=7500$ to $t=12500$ in Fig.~1a).
    We call the state in the quiescent phase laminar state $L$. 

    \begin{figure}[tbp]
        \centering
        \includegraphics[width=0.99\columnwidth]{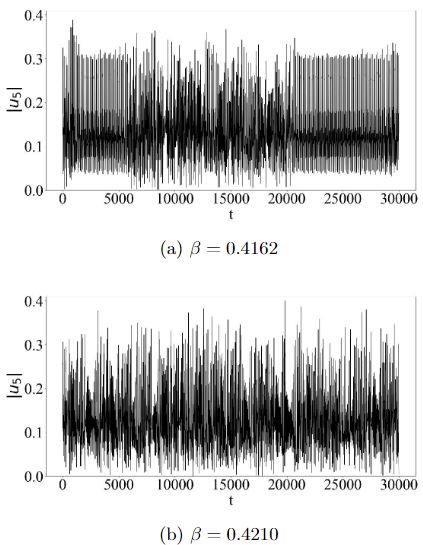}
        \caption{\textbf{Intermittent behavior of the shell model.} 
            The time evolution of $|u_5|$ for time length $T=3\times10^4$ is shown for (a) $\beta=0.4162$ and (b) $\beta=0.4210$. 
            Regarding $\beta=0.4162$, we can easily observe intermittent dynamics, and
            all variables $u_j$ of the shell model concurrently show such intermittency, although $u_j$ for larger $j$ tends to be more intermittent. At $\beta=0.4210$, we hardly observe the laminar state.
        }
        \label{fig:SM:intermittency}
    \end{figure}
    
\begin{figure}[tbp]
        \centering
        \includegraphics[width=0.99\columnwidth]{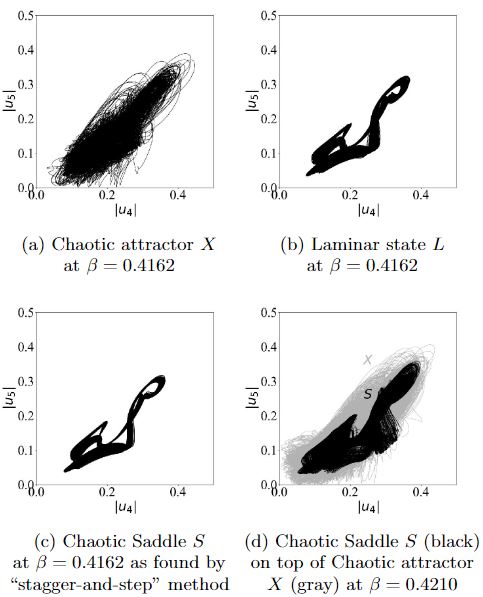}
        \caption{
            \textbf{Comparison of sets for the shell model after the crisis} projected onto $(|u_4|,|u_5|)$ space. 
            The panel~(a),~(b),~(c) show the chaotic attractor $X$, the laminar states $L$, and the chaotic saddle $S$ respectively at $\beta=0.4162$.
            The panel~(d) shows a chaotic saddle $S$ at $\beta=0.4210$, where laminar states are unobservable by examining the time series as Fig.~1b. The gray-colored trajectory is $X$ at $\beta=0.4210$.
        }
        \label{fig:SM:objects}
    \end{figure}

    We confirm that the laminar state corresponds to the chaotic saddle (Fig.~\ref{fig:SM:objects}) by employing stagger-and-step method. We obtain the chaotic saddle $S$ at $\beta=0.4162$ (Fig.~2c). 
    We also construct $L$ by isolating segments of the quiescent phase from $X$ through a process of ``trimming'' (Fig.~2b). 
    See Appendix I for the details of the ``trimming.'' 
    Both sets are almost identical. 
    Furthermore, in contrast to $L$, the size of $S$ remains unchanged over the 5000 time integration.
    
    The chaotic saddle $S$ is also observed at $\beta = 0.4210$ (Fig.~2d), far from $\beta^*$, where the laminar state is almost unobservable (Fig.~1b). This implies the existence of chaotic saddles even in regions far from the critical parameters. Furthermore, it shows that intermittency appears robustly in developed turbulence. 
    
    We calculated finite-time Lyapunov exponents along an orbit wandering on a chaotic saddle $S$. The result shows that $S$ contains periodic orbits with $1$--$6$ unstable directions, implying that $S$ is hetero-chaotic~\cite{Saiki2018}, and the intermittency is caused by such hetero-chaos structure of transverse stability to the laminar state. In coupled identical oscillators system, hetero-chaos structure of transverse stability is the key to understanding blowout bifurcation~\cite{nagai1997}. 

\begin{figure}[tbp]
        \centering
            \includegraphics[width=0.99\columnwidth]{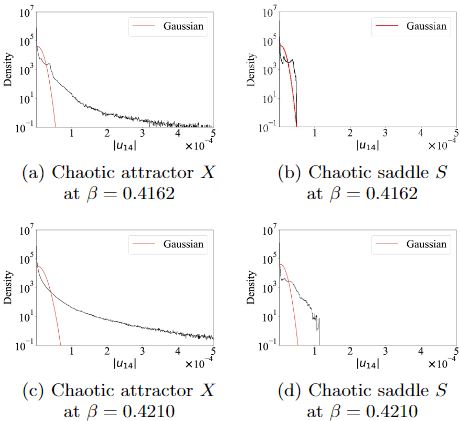}
        \caption{\textbf{Comparison of the distributions of $|u_{14}|$ with the Gaussian distribution.}
        Each subfigure shows the density distribution of $u_{14}$'s amplitude and the Gaussian distribution with equal mean and variance for each distribution.
        The left panels (a), (c) illustrate the distribution on the attractor $X$, that is, laminar and bursty states are included, and the right panels (b), (d) illustrate that on the chaotic saddle $S$.
        Regarding the chaotic saddle $S$, the distributions are close to the Gaussian distribution, whereas $X$ has a long tail. 
        This indicates that flow on the chaotic saddle corresponds to the laminar flow.
        }
        \label{fig:SM:distribution}
    \end{figure}

    Additionally, the chaotic saddle $S$ corresponds to the actual laminar flow in fluid dynamics.
    Intermittency in fluids is characterized by non-Gaussian distributions of the longitudinal velocity difference, with long tails and a bump around zero.
    The appearance of violent fluctuations amidst periods of small, quiet fluctuations causes such bias.
    In this paper, we observed distribution in the dissipation range.
    We compared the distributions of the amplitude of $u_{14}$ between the attractor and the chaotic saddle (Fig.~\ref{fig:SM:distribution}). 
    The distribution for the chaotic saddle is relatively close to the Gaussian distributions, while that for the attractor has a long tail.
    This indicates that the chaotic saddle $S$ corresponding to the laminar state $L$ also corresponds to the laminar flow. 
    
\section{III. {Phase synchronization in non-identical coupled chaotic model}} 
    This section presents the concept that a laminar chaotic saddle is a generalized form of a phase synchronized state. 
    For this purpose, we investigate the non-identical coupled R\"ossler oscillator.
    Regarding the coupling of identical systems, Lai \etal~\cite{lai1999b, lai1999a} showed the coexistence of heterogeneous periodic orbits with different transverse stability. Yanchuk \etal~\cite{yanchuk2001} reported that the blowout bifurcation occurs in coupled identical R\"ossler systems. 
    In this section, we investigate a low-dimensional coupled system composed of two different oscillators where the synchronized state is not trivial and unstable. 
    Using the model, we show that the phase synchronized states correspond to the chaotic saddle.
    
    The coupled R\"ossler model~\cite{pikovsky2001} is a simple model for observing phase synchronization between coupled chaotic systems.
    It is a 6-dimensional system of ordinary differential equations obtained by coupling two 3-dimensional R\"ossler models as follows:
    \begin{equation*}
  \begin{cases}
      \dot{x}_{1, 2} &= -\omega_{1, 2}y_{1, 2}-z_{1, 2}+\eps(x_{2, 1} -x_{1, 2}),\\
        \dot{y}_{1, 2} &= \omega_{1, 2}x_{1, 2}-ay_{1, 2},\\
        \dot{z}_{1, 2} &= f + z_{1, 2}(x_{1, 2} - c),
  \end{cases}
\end{equation*}
    where we adopted $a = 0.165,\ f=0.2,\ c=10,\ \omega_{1, 2}=\omega_0\pm \delta~(\omega_0=0.97,\ \delta=0.02),$ and various coupling strengths $\eps$. 
    The difference of $\omega_1$ and $\omega_2$ gives asymmetricity.
    This model shows phase synchronization between two oscillators when the coupling strength is sufficiently large~\cite{rosenblum1997, pazo2003}. 
    
    We define the phase of an oscillator $\phi_i (i=1,2)$ at time $t$ by the angle in the $xy$-plane as follows:
    $$\phi_i = \arctan\frac{y_i}{x_i}, \quad i=1, 2.$$
    A phase difference $\psi_{1,2}$ at time $t$ between the two oscillators is defined as follows:
    \begin{equation}
        \psi_{1, 2}(t) = |\hat{\phi_1}(t) - \hat{\phi}_2(t)|,
        \label{eq:phase_difference}
    \end{equation}
    where $\hat{\phi_i} (i=1,2)$ is a lift of $\phi_i$.
    We define that phases are synchronized if $\psi_{1,2}$ are locked to the vicinity of a certain value for a certain time. 
    See \cite{pazo2003} for further details. 

    As $\eps$ decreases from the parameter where the synchronized state is stable, the phase synchronization becomes unstable at the critical parameter $\eps^{\ast} \approx 0.0416$~\cite{crisis_coupled_rossler}. 
    Then, the attractor corresponding to the phase synchronized state becomes unstable, and a larger chaotic attractor $X$ corresponding to the intermittent state appears. The laminar state $L$ is equivalent to the phase synchronized set in the intermittent state (Fig.~\ref{fig:CR:phase-shift}).
    
    By employing the stagger-and-step method using the phase (see Appendix II for details), we identified the chaotic saddle $S$ at $\eps=0.039$ in the intermittent region (Fig.~\ref{fig:CR:objects}). 
    We compare the chaotic saddle $S$  (Fig.~5c) and the laminar state $L$  (Fig.~5b) identified from a trajectory on the attractor $X$ through the ``trimming.'' 
    Both sets are almost identical, and the chaotic saddle corresponds to the phase synchronization.  
    Similar to the shell model, $S$ is numerically verified to be an invariant set, whereas $L$ is not an invariant set. We also obtained chaotic saddle $S$ at $\eps = 0.025$ (Fig.~5d), far from $\eps^*$, where phase synchronization
    is unlikely to occur in a typical time series.

    As a further point of interest, by calculating finite-time Lyapunov exponents of the chaotic saddle $S$, we find that the number of positive finite-time Lyapunov exponents changes from $1$ to $3$ with time, implying that $S$ contains periodic orbits having $1$--$3$ unstable directions, and the hetero-chaotic property is expected to cause the intermittency.

    \begin{figure}[tbp]
        \centering        \includegraphics[width=0.99\columnwidth]{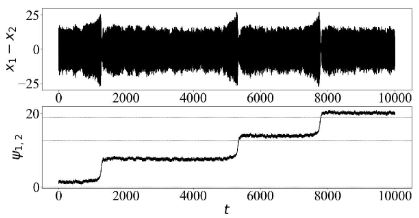}
        \caption{\textbf{Intermittency and phase synchronization of the coupled R\"ossler model for $\eps=0.039$.} 
        The top and bottom panels show the time evolution of $|x_1 - x_2|$ and phase difference $\psi_{1,2}$, respectively for the same time interval. $\psi_{1, 2}$ is defined in Eq.~\eqref{eq:phase_difference}. The slip of the phase difference coincides with the timing of the large fluctuation of the dynamics.}
        \label{fig:CR:phase-shift}
    \end{figure}

    \begin{figure}[tbp]
        \centering
            \includegraphics[width=0.99\columnwidth]{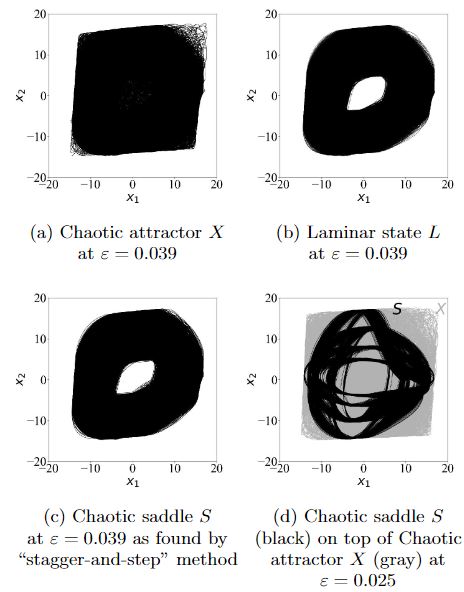}
        \caption{\textbf{Comparison of sets for the coupled R\"ossler model after the crisis.} The panels (a), (b), 
        and (c) show the chaotic attractor $X$ , the laminar set $L$ and the chaotic saddle $S$ respectively at $\eps=0.039$. Panel (d) shows the chaotic saddle $S$ on the chaotic attractor at $\eps = 0.025$, where phase synchronization is unlikely to occur in a typical time series. 
        The gray-colored trajectory is $X$ at $\eps=0.025$. 
        The laminar state $L$ is identified by trimming, and the chaotic saddle $S$ identified by the stagger and step method is a refined version of $L$.}
        \label{fig:CR:objects}
    \end{figure}

    We showed that there exists a hetero-chaotic saddle corresponding to the laminar state of both the fluid turbulence model and the coupled chaotic model. For the coupled model, the laminar state is equivalent to a phase synchronized state. Therefore, if we can define the phases, we expect the phases of the turbulence model (Eq. \ref{eq:shell-model}) to be also synchronized. 

    For each variable $u_j\ (j=1, 2,\cdots,N)$, we define the phase $\phi_j(t)$ by the argument Arg$(u_j(t))$ and phase difference $\psi_{ij}$ between $\phi_i$ and $\phi_j$ as the same as Eq.~\ref{eq:phase_difference}.
    We define that $u_i$ and $u_j$ are
    synchronized if $\psi_{ij}$ are locked to the vicinity of a certain value for a certain time. 
    As a result, we found that phases of the shell model partially synchronize in the laminar state. It has been confirmed that the phases of variables with subscripts that are congruent modulo 3 (\eg 8 and 11) are synchronized. For instance, Fig.~6a is the time evolution of $\phi_{8, 11}$ with $\beta = 0.4162$, the same trajectory as Fig.~1a, and $\psi_{8, 11}$ is locked within a narrow range during the laminar state. However,  fluctuations and jumps occur during a bursty state. Figure~6b illustrates the time evolution of $\psi_{8,11}$ on a chaotic saddle corresponding to the laminar state. It is observed that the phase difference remains locked within a narrow range at all.

    We characterized the phase synchronization by hetero-chaotic saddle in the coupled R\"ossler model. Compare with that, the phase synchronization in the shell model is not as clear as in the coupled R\"ossler model in some respects, including the partial synchronization. Even for this weak phase synchronization, the localized hetero-chaotic saddle $S$ is useful for the characterization.
    
    \begin{figure}
        \centering
        \includegraphics[width=0.99\columnwidth]{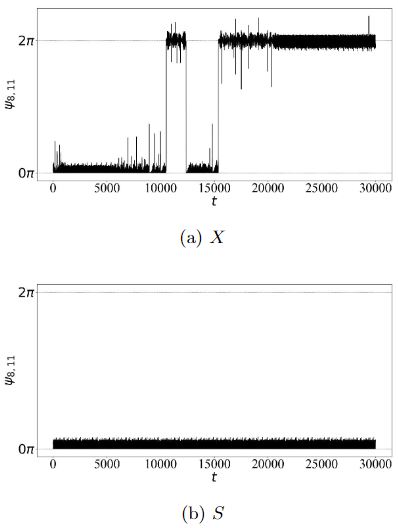}
        \caption{\textbf{Phase difference between $u_8$, $u_{11}$ in the shell model.} (a) is the time evolution of $\psi_{8, 11}$ with $\beta = 0.4162$, the same trajectory as Fig.1a
        and the phases of $u_8$ and $u_{11}$ are locked within a narrow range during the laminar state. However, during the bursty state, fluctuations and jumps occur. (b) illustrates the time evolution of the phases on a chaotic saddle $S$ corresponding to the laminar state. It is observed that the phase difference remains locked within a narrow range at all.}
        \label{fig:SM:phase}
    \end{figure}

\section{IV. Concluding remarks}
    \subsection{Summary}
        We investigated a chaotic saddle, namely 
        a non-attracting chaotic set that is a proper subset of a chaotic attractor on which the dynamics show intermittency.
        In a fluid turbulence model, the detected hetero-chaotic saddle within a chaotic attractor, 
        which is a refined version of the weakly chaotic (laminar) state, 
        corresponds to the laminar flow in developed turbulence. 
        In a phase synchronization model showing intermittency, the hetero-chaotic saddle corresponds to the phase synchronized state. This indicates phase synchronization between asymmetric oscillators can be characterized by hetero-chaotic saddle. 
        For each case, the chaotic saddle persists for a wide range of parameters. We also observed phase synchronization in the shell model. 
    \subsection{Discussion}
    In the study of (phase) synchronization, 
    a hetero-chaotic saddle with a different number of transversally unstable dimensions has sometimes been used to characterize a coherent state. However, such studies mainly examine symmetrically coupled identical oscillators or directionally coupled dynamics \cite{kapitaniak1999, yanchuk2001, glendinning2001}. 
    We investigated intermittency in asymmetric coupled dynamics, including model turbulence and coupled chaotic oscillators. 
    Characterizing the stability of the weakly chaotic coherent state is rather complicated, mainly because finding the transverse direction of the chaotic saddle 
    is difficult. Hence, the characterization of blowout bifurcation is not straightforward. 
    Our approach focuses on the appearance/disappearance of a chaotic saddle and thus gives new insight into characterizing the destabilization of coherent dynamics. 
    Furthermore, the existence of intermittency is often difficult to determine in the dynamics we consider. Using our idea, intermittency is judged by the existence of a hetero-chaotic saddle within a chaotic attractor.

    We confirmed the occurrence of phase synchronization in the shell model. This indicates that turbulence models can be generalized as coupled chaotic models~\cite{haugland}. 
    In high-dimensional systems, phases are not defined appropriately in general. 
    Even in such cases, a laminar hetero-chaotic saddle like cases in this paper can be used as a generalization of a phase synchronized state.

    In this paper, we consider the laminar chaotic saddles. It is anticipated that there also exist chaotic saddles corresponding to the bursty state. Rempel \etal~\cite{rempel2007a} identified both laminar and bursty chaotic saddles embedded in chaotic attractor, the latter of which are not localized in the attractor. See Figs. 6d and 7 in~\cite{rempel2007a}.
    
     Chaos control is usually aimed at 
        using feedback control to stabilize a periodic orbit in a chaotic attractor~\cite{ott1990}.
        We believe that this study opens the door to controlling weakly chaotic sets in a highly chaotic attractor. See~\cite{zambrano2008}.
        An example would be stabilizing intermittent laminar pipe flow using feedback control.

\section{Appendix}
    \subsection{I. Trimming}    
\begin{figure}[b]
        \centering
        \includegraphics[width=0.99\columnwidth]{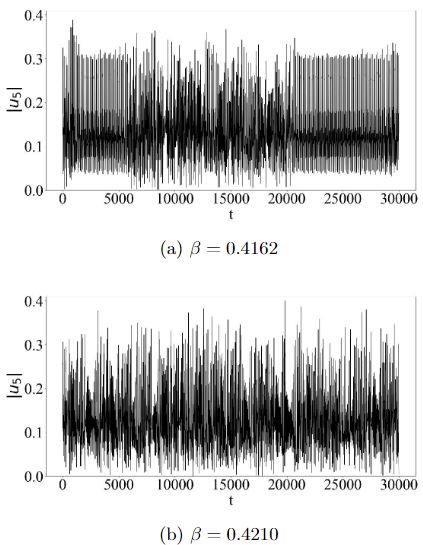}
        \caption{\textbf{Schematic picture of trimming.}
            For extracting the laminar state, we removed the beginning and last parts of each quiescent phase, aiming to obtain a set closer to the invariant set. 
            The top panel is a projection of a trajectory of the coupled R\"ossler model. 
            The bottom panel shows the time evolution of the phase difference $\psi_{1,2}$. 
            Each gray area in the top and bottom panels denotes the projection of a chaotic trajectory and phase difference showing intermittency, respectively. Each black area denotes those of the extracted weakly chaotic dynamics.
            }
        \label{fig:trimming}
    \end{figure}

    We identified and extracted laminar states from time series showing intermittency. When extracting a laminar state, we removed the transient to obtain the set closer to an invariant set. In this study, we removed the beginning and end of each extracted trajectory for a certain amount of time, which we call trimming.
    Regarding the coupled R\"ossler model, we extracted the trajectory that showed phase synchronization continuously for more than time length $T=500$, corresponding to 100 times the lowest period of periodic orbits. The left panel of Fig.~\ref{fig:trimming} shows the extracted trajectories. 
    Then, we removed the first and last time length $T = 250$ of each extracted phase synchronized part. The right panel of Fig.~\ref{fig:trimming} shows the result, and the removal of the transients can also be seen from the trajectory. 

    \subsection{II. Phase introduced stagger-and-step}
        In the traditional stagger-and-step method proposed by Sweet \etal~\cite{sweet2001}, trajectories that remain in a particular region for a long time are considered to form an approximation of an invariant set.
        In this paper, we propose an alternative approach in which trajectories approximately synchronous for a long time are considered close to invariant sets. For the coupled R\"ossler model, laminar states correspond to the phase synchronized states. Thus, phase synchronization can be a sufficient criterion for identifying chaotic saddles; it is better than the conventional stagger-and-step method in certain respects.
        
        The conventional stagger-and-step method relies on prior knowledge of the general shape of chaotic saddles. The region that sufficiently covers the shape robustly must be defined. The region choice could potentially increase the computational load in high-dimensional dynamical systems. 
    
        In contrast, the phase synchronization method does not require prior knowledge of the shape of the chaotic saddle.
        We can identify the chaotic saddles of high-dimensional dynamical systems with relatively low computational costs by examining phase differences between selected variables.

\section{Acknowledgements} 
YS was supported by JSPS KAKENHI Grant Nos. 19KK0067, 21K18584, 23H04465, and 24K00537 and JSPS Bilateral Open Partnership Joint Research Projects JPJSBP 120229913. 
The computation was performed using the JHPCN (jh230028 and jh240051).
  
%


\begin{thebibliography}{39}
\bibitem [{hyper-chaos()}]{hyper-chaos}%
  \BibitemOpen
  \bibinfo {note} {Hyperchaos is defined by the presence of multiple positive
  Lyapunov exponents, indicating the existence of at least two different
  unstable directions. Unlike hetero-chaos, hyperchaos does not involve the
  diversity of unstable periodic orbits}\BibitemShut {NoStop}%
\bibitem [{crisis_coupled_rossler()}]{crisis_coupled_rossler}%
  \BibitemOpen
  \bibinfo {note} {The crisis is thought to be accompanied by the stability
  change in the ``transverse'' direction. This differs from the case for an
  attractor-merging crisis~\cite{saiki2011}.}\BibitemShut {Stop}%
\end{thebibliography}
\end{document}